\documentclass[preprint,12pt]{elsarticle}

\usepackage{amssymb}
\usepackage{color}

\journal{}

\begin{document}

\begin{frontmatter}



\title{Multiple learning mechanisms promote cooperation in public goods games with project selection}


\author{Li-Xin Zhong$^a$}\ead{zlxxwj@163.com}
\author {Wen-Juan Xu$^b$}
\author {Rong-Da Chen$^a$}
\author {Yun-Xin He$^a$}
\author {Tian Qiu$^c$}
\author {Fei Ren$^d$}
\author {Yong-Dong Shi$^e$}
\author{Chen-Yang Zhong$^f$}

\address[label1]{School of Finance and Coordinated Innovation Center of Wealth Management and Quantitative Investment, Zhejiang University of Finance and Economics, Hangzhou, 310018, China}
\address[label2]{School of Law, Zhejiang University of Finance and Economics, Hangzhou, 310018, China}
\address[label3]{School of Information Engineering, Nanchang Hangkong University, Nanchang, 330063, China}
\address[label4]{School of Business, East China University of Science and Technology, Shanghai 200237, China}
\address[label5]{School of Applied Finance and Behavioral Science, Dongbei University of Finance and Economics, Dalian, 116025, China}
\address[label6]{Department of Statistics, Stanford University, Stanford, CA 94305-4065, USA}

\begin{abstract}
How evolution favors cooperation is a fundamental issue in social and economic systems. In the business world, actively selecting a suitable project is usually helpful for a businessman to be in an advantageous position. By incorporating project selection mechanism into the threshold public goods game, we have investigated the coupling effect of mutation and imitation in updating one's preferred project scale on the evolution of cooperation. Compared with the situation where there is no project selection mechanism, the existence of project selection may suppress or promote cooperation depending upon different updating rules. There exists a critical ratio of the imitators in the population, below which cooperation is suppressed while above which cooperation is promoted. With the coevolving of individual strategies and preferred project scales, a higher level of cooperation corresponds to a larger average value of the preferred project scales. A theoretical analysis indicates that, as most of the individuals are mutants, the coevolving process is governed by the mutation process, which leads to a smaller average value of the preferred project scales and a lower level of cooperation. As most of the individuals are imitators, the coevolving process is governed by the coupling of the mutation and imitation processes, which leads to a larger average value of the preferred project scales and a higher level of cooperation. As all the individuals are imitators, the coevolving process is governed by the imitation process, which leads to an intermediate average value of the preferred project scales and an intermediate level of cooperation.
\end{abstract}

\begin{keyword}
public goods game \sep project selection \sep multiple learning mechanism \sep individual preference

\end{keyword}

\end{frontmatter}


\section{Introduction}
\label{sec:introduction}
The emergence and maintenance of cooperation among unrelated and selfish individuals is a fundamental and fascinating issue in
economical, social and biological systems\cite{perc1,sornette1,yukalov1,schweitzer1, mcavoy1,wardil1,qiu1,hadzibeganovic1,hadzibeganovic2}. In the last two decades, based on the models of the prisoner's dilemma, the snowdrift game and the public goods game, tremendous efforts have been dedicated to the exploration of the conditions for the occurrence of  mutually beneficial interactions, among which the mechanisms of connectivity, reputation, penalization and non-participation have been extensively investigated\cite{zhong1,chotibut1,ginsberg1,szolnoki1,darcet1,szolnoki2,helbing1,helbing2,hauert1}.

The public goods game describes such a scenario where each individual in a group has the opportunity to make a contribution to a common pool or not\cite{liu1,chen1,chen2,szolnoki3,zhong2,perc2,meloni1,jwang1,gomez-gardenes1}. All the contributions in the common pool are multiplied by a factor $r$ and divided equally among all the group members irrespective of whether an individual has contributed to the common pool or not. Because only the cooperators bear the burden of contributions, the replicator dynamics in such a game model is in favor of the defectors and the cooperators eventually become extinct.

More recently, the effects of coevolving of individual strategies and network structures on the evolution of cooperation have received special attention \cite{zhang1,lin1,ebel1,zimmermann1,bekiros1,zwang1,allen1,takesue1,perc3,traulsen1}. The relinking preferences have been found to play an important role in the evolution of individual strategies and network structures\cite{pacheco1,santos1,szolnokis4,szolnokis5}. In addition to that, the effects of heterogeneous preferences for strategy updating on the evolution of cooperation have been studied\cite{danku1,szolnoki6,amaral1}. It has been found that the multiple learning mechanisms, like innovation and imitation in updating one's strategy, can help sustain cooperation in the situations where the defectors would invade with only one updating rule.

In real society, people are not always passively participating in an activity. For example,in project bidding, people tend to choose the projects that fit their personal capabilities and available resources. Faced with a variety of preys that might be captured, the hunters have to make a decision on whether to hunt a big or a small prey. The above scenario is related to the problems of start-up costs\cite{takano1,chen3,szolnoki7}. Although similar problems like the role of critical mass in the evolution of cooperation have been discussed\cite{huang1,sigmund1,stollmeier1},  how an active selection affects the evolution of cooperation is still an open problem.

Inspired by the studies in the fields of coevolving dynamics and heterogeneous preferences\cite{pacheco1,santos1,szolnokis4,szolnokis5,danku1,szolnoki6,amaral1}, in the present work, we incorporate project selection mechanism into the threshold public goods game\cite{zhong3}. Both the coupling effect of individual strategies and preferred project scales and the coupling effect of mutation and imitation in updating one's preferred project scale on the evolution of cooperation have been investigated. The main motivations of the present work are as follows.

The first motivation of the present work is to investigate how the project selection mechanism affects the cooperation dynamics. In the original public goods game, although the project selection mechanism has not been discussed, there really exists an adjustable project in the competing process, which depends on the number of cooperators in the group. For example, if there are $n_C$ cooperators in the group and the contribution of each cooperator is I, the scale of the project selected by the group should be $S=n_CI$. Therefore, the project can always be finished in the original public goods game. However, in the real world, the project may be given before the number of cooperators has been known. For example, after we have got a project in a bid, we may finally give it up because we can not find enough cooperators. The group members get nothing in such a situation. If there are more cooperators than the needs to finish the project, each cooperator does not need to make his maximal contribution. Compared with the situation where there are just enough cooperators in the group, the benefit of each individual has no change while the cost of each cooperator decreases, which is similar to the N-person snowdrift game introduced in reference\cite{zheng1}. Therefore, the present model is a generalization of the original public goods game and the original N-person snowdrift game. As the number of cooperators is known before the project is selected, the present model restores to the original public goods game.

The second motivation of the present work is to investigate how the updating rules affect the coevolving of individual strategies and preferred project scales. In the original public goods game, the coevolving of individual strategies and network structures has been discussed. However, in the real world, there exist a variety of individual preferences\cite{bekiros1,szolnokis4,szolnokis5}, like the preferred group size and the preferred project scale, which may also evolve with time. The coevolving dynamics of individual strategies and preferred group sizes has been discussed in references\cite{szolnokis4,zhong3}. The coevolving dynamics of individual strategies and preferred project scales is still an open issue which needs to be investigated in depth. As to the effects of multiple updating rules on the evolution of cooperation, most researches have discussed the effects of multiple updating rules on the evolution of individual strategies. In the present work, we have discussed the effects of multiple updating rules on the coevolving of individual strategies and preferred project scales.

Therefore, in the present work, we need to answer the following two questions: can the coevolving of individual strategies and preferred project scales affect cooperation dynamics? If yes, how do the multiple updating rules affect the levels of cooperation? The main findings are as follows.

(1)The project selection mechanism is not always beneficial for cooperation. Compared with the situation where there is no project selection mechanism, the existence of imitation in project selection is beneficial for cooperation while the existence of mutation in project selection is detrimental to cooperation.

(2)The effects of multiple updating rules on the evolution of cooperation depend upon the ratio of imitators in the population. There exists a critical ratio of the imitators in the population, below which cooperation is suppressed while above which cooperation is promoted. A higher level of cooperation corresponds to a larger average value of the preferred project scales.

(3)A theoretical analysis indicates that a higher level of cooperation results from the optimum matching between the group size and the project scale. The multiple updating rules are more advantageous for an individual to find the most appropriate project, which helps reach the highest level of cooperation.

\section{The model}
\label{sec:model}
In the original public goods game, the competing individuals are well-mixed or arranged on a network\cite{szabo1,hauert2}. In a well-mixed population, each individual can interact with all the other individuals, which corresponds to the real world situation where the group members are chosen randomly for the need to accomplish a specific task, such as a one-off project and a temporary assignment. The group will be disbanded after the project or the assignment has been finished. In a networked population, each individual can only interact with his immediate neighbors with direct connections, which corresponds to the real world situation where the group members are relatively stable, such as family members and enterprise personnel. Some studies have shown that the network structure can effectively influence the evolving process in the public goods game\cite{szabo2,szolnoki8,wang1,shapiro1}. In order to focus on the impact of project selection mechanism and refrain from the coupling effects of project selection and relinking on cooperation, we only consider the well-mixed case in the present model.

In the public goods game without project selection mechanism\cite{hauert2,doebeli1}, a large population $N$ consists of cooperators and defectors. From time to time, $n$ randomly chosen individuals engage in the public goods game. If the group of $n$ individuals of the sample consist of $n_C$ cooperators and $n-n_C$ defectors and a cooperator's investment is $I$, the payoff of a defector is $P_D=rIn_C/n$ and the payoff of a cooperator is $P_C=P_D - I$. In the process of strategy updating, an individual i firstly selects an individual j randomly. If individual i's strategy is different from individual j's, individual i adopts individual j's strategy with probability $\omega$,

\begin{equation}
\label{eq.2}
\omega_{i\gets j}=\frac{1}{1+e^{(P_i-P_j+\tau)/\kappa}},
\end{equation}
in which $P_i$ is individual i's payoff, $P_j$ is individual j's payoff, $\tau=\kappa=0.1$ represents an occasional drift of the strategy because of the environmental impact. Or else, individual i keeps his strategy.

In the public goods game with project selection mechanism, before an individual i's strategy is updated, he firstly has to choose his group members and make a judgement about whether he should update his preferred project scale $S_i$ or not. If there are enough cooperators in the group, that is, $n_CC^{max}_I\ge S_i$, in which $C^{max}_I$ is a cooperator's maximal contribution, the project $S_i$ can be finished and he will get a cooperator's payoff $P_C=\frac{rS_i}{n}-\frac{S_i}{n_C}$ or a defector's payoff $P_D=\frac{rS_i}{n}$. Or else, the project can not be finished and he will get a cooperator's or a defector's payoff $P_C=P_D=0$. If his payoff $P_i$ is less than the average payoff $\bar P$ of the population, he modifies his preferred project scale. Or else, he keeps his preferred project scale. Inspired by the work done in references\cite{danku1,szolnoki6,amaral1}, concerning the update of one's preferred project scale, we incorporate two kinds of individuals, called $mutants$ and $imitators$ respectively, into the present model. A mutant i modifies his project $S^0_i$ in the following way: he randomly chooses a new project $S_i$ within the range of $S_i\in[S^0_i-\frac{R}{2}, S^0_i+\frac{R}{2}]$. If $S_{min}\le S_i\le S_{max}$, $S^0_i=S_i$. If $S_i<S_{min}$, $S^0_i=S_{min}$. If $S_i>S_{max}$, $S^0_i=S_{max}$. An imitator $i$ modifies his project $S^0_{i}$ in the following way: he firstly finds an individual j with the highest payoff $P_{j}$. Then he replaces his project with individual j's project, $S^0_{i}=S^0_{j}$. The above project selection mechanism is similar to the real world situation where a reasonable investor has been trying to find a more profitable project\cite{bekiros2,bekiros3,andreasson1,nax1}. If his project return is lower than the industry average return, he  will reselect his investment project. The multiple updating rules in project selection represent the heterogeneous personalities.

After all the individuals have got their preferred project scales, each individual will make a decision on whether he should update his strategy or not. An individual's payoff is obtained as that in the project updating process and the update of an individual's strategy is as that in the original public goods game. The distributions of the preferred project scales and the frequencies of cooperators and defectors evolve over time.

Different from the situation in the original public goods game where the total investment depends on the composition of the group\cite{hauert2,doebeli1}, in the present model, the total investment is decided before the group members are selected and does not depend on the composition of the group. Such a difference may lead to the occurrence of the following two scenarios which do not occur in the original public goods game. One scenario is that the cooperators and defectors may get nothing even if there are cooperators in the randomly chosen group. Another scenario is that a cooperator may only invest a part of his money into the public goods. Therefore, if the project is chosen according to the group composition, the present model restore to the original model.

\section{Simulation results and discussions}

\begin{figure}
\centering
\includegraphics[width=5cm]{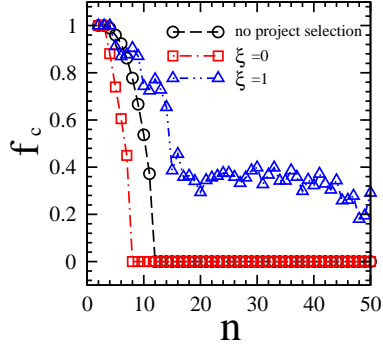}
\caption{\label{fig:epsart}The frequency of cooperators $f_C$ as a function of the group size $n$ for the cases where there is no project selection mechanism (circles), all the individuals are mutants (squares), all the individuals are imitators (triangles). Other parameters are: the total population $N=5000$, an individual's ability (maximal contribution) $C^{max}_I$=2, the multiplication factor $r=5$, the minimum project scale $S_{min}$=1, the maximum project scale $S_{max}$=$nC^{max}_I+1$. Final results are averaged over 10 runs and 1000 time steps with 5000 relaxation time in each run.}
\end{figure}

The first question we asked is that, compared with the situation where there is no project selection mechanism, the existence of project selection mechanism can promote or impede cooperation.

Figure 1 shows the frequency of cooperators $f_C$ as a function of the group size n in the situation where there is no project selection mechanism and the situations where there is only one updating rule in project selection. As can be seen from the results in Figure 1, compared with the situation where there is no project selection mechanism, the existence of project selection may be beneficial for or detrimental to cooperation, which depends on the updating rule in project selection. If all the individuals are mutants, the cooperation is impeded. If all the individuals are imitators, the cooperation is promoted.

\begin{figure}
\centering
\includegraphics[width=5cm]{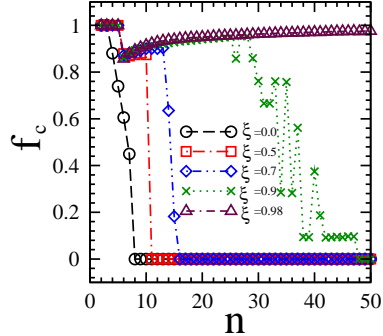}
\caption{\label{fig:epsart}The frequency of cooperators $f_C$ as a function of the group size $n$ for the ratio of imitators $\xi$=0 (circles), 0.5 (squares), 0.7 (diamonds), 0.9 (crosses), 0.98 (triangles). Other parameters are: the total population $N=5000$, an individual's ability (maximal contribution) $C^{max}_I$=2, the multiplication factor $r=5$, the minimum project scale $S_{min}$=1, the maximum project scale $S_{max}$=$nC^{max}_I+1$. Final results are averaged over 10 runs and 1000 time steps with 5000 relaxation time in each run.}
\end{figure}

The second question we asked is that, compared with the situation where there is only one updating rule in project selection, multiple updating rules can promote or impede cooperation.

Figure 2 shows the frequency of cooperators $f_C$ as a function of the group size n in the situations where there are two kinds of individuals, mutants and imitators, in project selection. As can be seen from the results in Figure 2, compared with the situation where all the individuals are mutants or imitators, the coexistence of mutants and imitators is beneficial for cooperation. Most imitators accompanied by a few of mutants can lead to the highest level of cooperation.

\begin{figure}
\centering
\includegraphics[width=5cm]{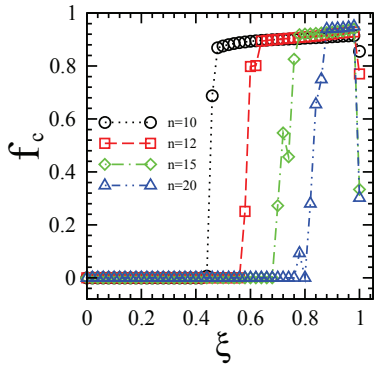}
\caption{\label{fig:epsart}The frequency of cooperators $f_C$ as a function of the ratio of imitators $\xi$ for the group size $n=10$ (circles), 12 (squares), 15 (diamonds), 20 (triangles). Other parameters are: the total population $N=5000$, an individual's ability (maximal contribution) $C^{max}_I$=2, the multiplication factor $r=5$, the minimum project scale $S_{min}$=1, the maximum project scale $S_{max}$=$nC^{max}_I+1$. Final results are averaged over 10 runs and 1000 time steps with 5000 relaxation time in each run.}
\end{figure}

The third question we asked is what the relationship between the frequency of cooperators and the ratio of imitators is.

In Figure 3 we plot the frequencies of cooperators as a function of the ratio of imitators in the population for different n. As can be seen from the results in Figure 3, there exists a transition point of the ratio of imitators in the population, below which the cooperators are doomed and above which a higher level of cooperation is reached. Except the situation where all the individuals are imitators, the frequency of cooperators increases with the rise of the ratio of imitators in the population. The transition point increases with the rise of the group size.

\begin{figure}
\centering
\includegraphics[width=12cm]{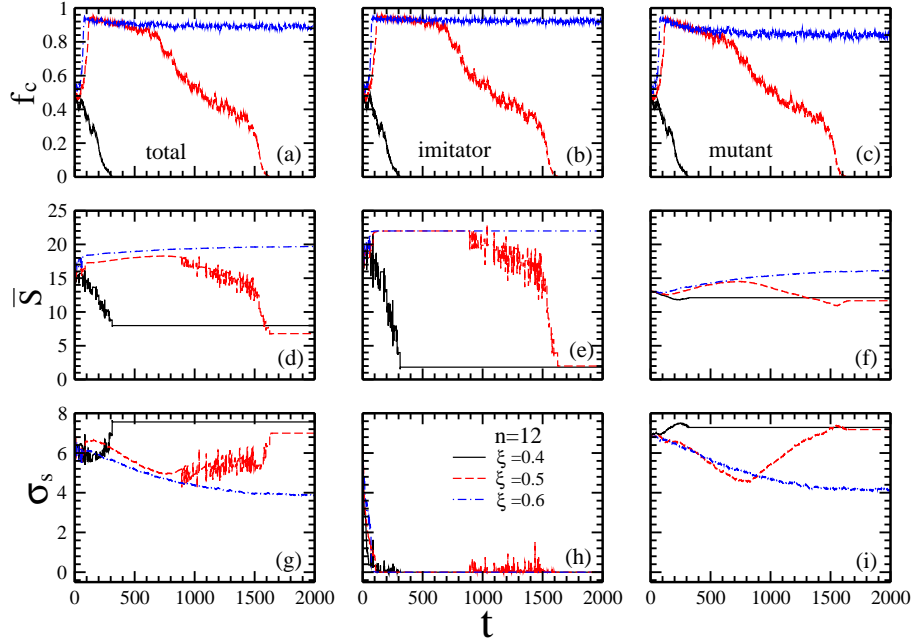}
\caption{\label{fig:epsart} The time evolution of the frequency of cooperators $f_C$, the average value of the preferred project scales $\bar S$, the standard deviation of the preferred project scales $\sigma_s$ for all the population (a, d, g), the imitators (b, e, h), the mutants (c, f, i) respectively. The ratio of imitators is $\xi=$ 0.4 (solid lines), 0.5 (dashed lines), 0.6 (dashed dotted lines). Other parameters are: the total population $N=5000$, the multiplication factor $r=5$, the group size $n=12$, an individual's ability (maximal contribution) $C^{max}_I$=2, the minimum project scale $S_{min}$=1, the maximum project scale $S_{max}$=$nC^{max}_I+1$.}
\end{figure}

The fourth question we asked is what the time-dependent behaviors of the frequency of cooperators and the distribution of the preferred project scales are.

In Figure 4 we plot the dynamic behaviors of the frequency of cooperators, the average value and the standard deviation of the preferred project scales. Figure 4 (a), (d) and (g) show that, when there is only a small ratio of imitators in the population, the frequency of cooperators decreases to zero rapidly over time. Accordingly, the average value of the preferred project scales decreases while the standard deviation of the preferred project scales increases rapidly over time. Such results indicate that, as the frequency of cooperators decreases, not all the individuals have chosen the small projects. An increase in the standard deviation of the preferred project scales represents an increase in the heterogeneity of the preferred project scales, which is detrimental to cooperation. When there is a moderate ratio of imitators in the population, the frequency of cooperators firstly increases quickly and then decreases continuously to zero over time. Accordingly, the average value of the preferred project scales firstly increases and then decreases continuously over time. The standard deviation of the preferred project scales firstly decreases and then increases continuously over time. When there is a large ratio of imitators in the population, the frequency of cooperators firstly increases quickly and then keeps a higher level over time. Accordingly, the average value of the preferred project scales firstly increases and then keeps a higher level over time. The standard deviation of the preferred project scales firstly decreases and then keeps a lower level over time.

In order to find out the differences between the imitators' and the mutants' behaviors, we plot the imitators' dynamic behaviors in Figure 4 (b), (e), (h) and the mutants' dynamic behaviors in Figure 4 (c), (f), (i) respectively. For the imitators, a lower cooperation corresponds to a smaller average value and a smaller standard deviation of the preferred project scales. A higher cooperation corresponds to a larger average value and a smaller standard deviation of the preferred project scales. For the mutants, a lower cooperation corresponds to a smaller average value and a larger standard deviation of the preferred project scales. A higher cooperation corresponds to a larger average value and a smaller standard deviation of the preferred project scales. The changes in the preferred project scales lag behind the changes in the frequency of cooperators. The simulation results tell us that a higher level of cooperation corresponds to the situation where nearly all the individuals adopt a larger project.

\begin{figure}
\centering
\includegraphics[width=10cm]{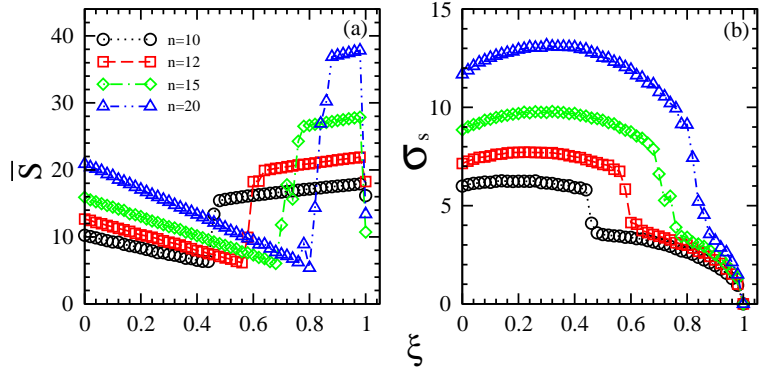}
\caption{\label{fig:epsart}(a) The average value $\bar S$ and (b) the standard deviation $\sigma_s$ of the preferred project scales as a function of the ratio $\xi$ of the imitators in the population for the group size $n=$10 (circles), 12 (squares), 15 (diamonds), 20 (triangles). Other parameters are: the total population $N=5000$, an individual's ability (maximal contribution) $C^{max}_I$=2, the multiplication factor $r=5$, the minimum project scale $S_{min}$=1, the maximum project scale $S_{max}$=$nC^{max}_I+1$. Final results are averaged over 10 runs and 1000 time steps with 5000 relaxation time in each run.}
\end{figure}

The fifth question we asked is what the relationship between the distribution of the preferred project scales and the ratio of imitators in the population is.

In Figure 5 we plot the average value and the standard deviation of the preferred project scales as a function of the ratio of imitators for different n. As can be seen from the results in Figure 5 (a), there exists a transition point of the ratio of imitators, below which the average value of the preferred project scales decreases with the rise of the ratio of imitators. Near the transition point, the average value of the preferred project scales increases quickly with the rise of the ratio of imitators. Above the transition point, the average value of the preferred project scales increases slowly with the rise of the ratio of imitators. As all the individuals are imitators, the average value of the preferred project scales has a large drop. As can be seen from the results in Figure 5 (b), there also exists a transition point of the ratio of imitators, below which the standard deviation of the preferred project scales firstly has a slow increase and then has a slow decrease with the rise of the ratio of imitators. Near the transition point, the standard deviation of the preferred project scales decreases quickly with the rise of the ratio of imitators. Above the transition point, the standard deviation of the preferred project scales decreases slowly with the rise of the ratio of imitators. For a given n, a larger average value of the preferred project scales corresponds to a smaller standard deviation of the preferred project scales.

Comparing the results in Figure 5 with the results in Figure 3, we find that the transition points in Figure 5 corresponds to the transition points in Figure 3, which indicates that the frequency of cooperators is closely related to the distribution of the preferred project scales.

\begin{figure}
\centering
\includegraphics[width=10cm]{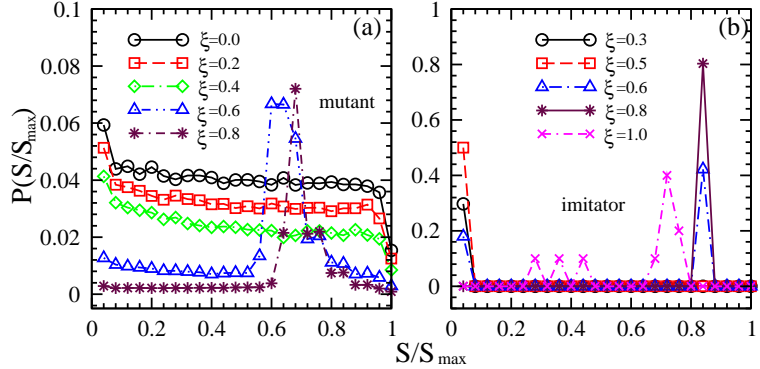}
\caption{\label{fig:epsart} The distribution of the preferred project scales (a) selected by the mutants for the ratio of imitators $\xi$=0 (circles), 0.2 (squares), 0.4(diamonds), 0.6(triangles), 0.8(stars); (b) selected by the imitators for the ratio of imitators $\xi$=0.3 (circles), 0.5 (squares), 0.6(triangles), 0.8(stars), 1(crosses). Other parameters are: the total population $N=5000$, the multiplication factor $r=5$, an individual's ability (maximal contribution) $C^{max}_I$=2, the minimum project scale $S_{min}$=1, the maximum project scale $S_{max}$=$nC^{max}_I+1$, the group size $n=12$. Final results are averaged over 10 runs and 1000 time steps with 5000 relaxation time in each run.}
\end{figure}

The sixth question we asked is, for a given ratio of imitators in the population, what kind of distribution the selected project scales meet.

In Figure 6 we plot the distributions of the preferred project scales selected by the mutants and the imitators respectively. As can be seen in Figure 6, as nearly all the individuals are mutants, the distribution of the project scales selected by the mutants is like an exponential distribution while the distribution of the project scales selected by the imitators is like a U-distribution. As nearly all the individuals are imitators, the distribution of the project scales selected by the mutants is like a Poisson distribution while the distribution of the project scales selected by the imitators is like a delta-distribution. Comparing the results in Figure 6 with the results in Figure 3, we find that, as nearly all the individuals select large projects, the cooperation is relatively high.

We conclude that the Poisson distribution or the delta distribution with a large average value of the preferred project scales is beneficial for cooperation.

\section{Theoretical analysis}
\label{sec:analysis}
\subsection{\label{subsec:levelA} replicator dynamics of the project scales selected by mutants and imitators}

Firstly, consider the case where all the individuals are mutants. The individuals can be divided into three categories according to  their preferred project scales: the number of individuals with a small project ($N_s$), the number of individuals with an intermediate project ($N_m$) and the number of individuals with a large project ($N_l$). Initially, $N^0_s=N^0_m=N^0_l$. As the frequency of cooperators is quite small, i.e. $f_C=\frac{1}{N}$, given a group size n, the payoff of the cooperator with a small project may be less than 0, that is, $P_C=\frac{rC_I}{n}-C_I<0$ on condition that $C_I\ge S$ and $r<n$. The payoff of the cooperator with an intermediate or a large project may be equal to 0, that is, $P_C=0$ on condition that $C_I<S$. Therefore, the cooperator with a small project is quite possible to update his project and the numbers of the individuals with different projects become $N_s=N^0_s-1$, $N_m=N^0_m+1$, $N_l=N^0_l$ . Given a group size n, the payoff of the defectors with a small project may be greater than 0, that is, $P_D=\frac{rC_I}{n}>0$ on condition that $C_I\ge S$. The payoff of the defector with an intermediate or a large project may be equal to 0, that is, $P_D=0$ on condition that $C_I<S$. The defectors with an intermediate or a large project is quite possible to update his project and the numbers of the individuals with different projects become $N_s=N^0_s-1+\Delta N$, $N_m=N^0_m+1$, $N_l=N^0_l-\Delta N$. Because there are more defectors than cooperators in the population, $f_D>>f_C$ and $\Delta N>>1$, the updating of the projects is quite possible to result in an increase in $N_s$ and a decrease in $N_l$, which is in accordance with the simulation result in fig. 6 (a) for $\xi=0$.

Secondly, consider the case where all the individuals are imitators. Initially, the preferred project scales are uniformly distributed, $N^0_s=N^0_m=N^0_l$. The ratios of cooperators in a group can be divided into three categories: small ($\frac{n^s_C}{n}$), intermediate ($\frac{n^m_C}{n}$), large ($\frac{n^l_C}{n}$). As the ratios of cooperators $\frac{n^s_C}{n}$, $\frac{n^m_C}{n}$ and $\frac{n^l_C}{n}$ coexist, for the cooperators, the individual with a large project and the largest ratio of cooperators in the group may have the highest payoff, i.e. $\frac{n^l_C}{n}=1$ and $S=nC_I$, the highest payoff of cooperators is $P_C=\frac{rS}{n}-\frac{S}{n_C}=(r-1)C_I$. For the defectors, the individual with a large project and the second largest ratio of cooperators in the group may have the highest payoff, i.e. $\frac{n^l_C}{n}=\frac{n-1}{n}$ and $S=(n-1)C_I$, the highest payoff of defectors is $P_D=\frac{rS}{n}=\frac{r(n-1)C_I}{n}$. With the imitation learning mechanism, all the individuals will adopt a large project. As the ratios of cooperators $\frac{n^s_C}{n}$ and $\frac{n^m_C}{n}$ coexist and $\frac{n^l_C}{n}=0$, for both the cooperators and the defectors, the individuals with an intermediate project may have the highest payoff, i.e. $\frac{n^m_C}{n}=\frac{1}{2}+\frac{1}{n}$ and $S=(\frac{n}{2}+1)C_I$ for a cooperator and $\frac{n^m_C}{n}=\frac{1}{2}$ and $S=\frac{nC_I}{2}$ for a defector, the highest payoff of cooperators is $P_C=\frac{rS}{n}-\frac{S}{n_C}=(\frac{r}{2}+\frac{r}{n}-1)C_I$ and the highest payoff of defectors is $P_D=\frac{rS}{n}=\frac{rC_I}{2}$. With the imitation learning mechanism, all the individuals will adopt an intermediate project. As only the ratio of cooperators $\frac{n^s_C}{n}$ exists and $\frac{n^m_C}{n}=\frac{n^l_C}{n}=0$, the defector with a small project may have the highest payoff, i.e. $\frac{n^s_C}{n}=\frac{1}{n}$ and $S=C_I$, the highest payoff of defectors is $P_D=\frac{rS}{n}=\frac{rC_I}{n}$. With the imitation learning mechanism, all the individuals will adopt a small project. For the initial condition $f_C\sim0.5$, it is quite possible that $\frac{n^s_C}{n}$ and $\frac{n^m_C}{n}$ coexist or $\frac{n^s_C}{n}$, $\frac{n^m_C}{n}$ and $\frac{n^l_C}{n}$ coexist. Therefore, the system may evolve to the state where the individuals adopt an intermediate project or a large project, which is in accordance with the simulation results in fig. 6 (b) for $\xi=1$.

Finally, consider the case where the mutants and the imitators coexist. For a small $\xi$, which corresponds to the situation where most individuals are mutants, the replicator dynamics of the project scales selected by the mutants is similar to the situation where all the individuals are mutants while the replicator dynamics of the project scales selected by the imitators is similar to the situation where all the individuals are imitators and only $\frac{n^s_C}{n}$ exists. In such a case, most individuals tend to adopt small projects. For a large $\xi$, which corresponds to the situation where most individuals are imitators, the replicator dynamics of the project scales selected by the imitators is similar to the situation where $\frac{n^s_C}{n}$, $\frac{n^m_C}{n}$ and $\frac{n^l_C}{n}$ coexist. For the mutants, because there are a large number of cooperators in the population, the individuals with a large project may have a higher payoff than the individuals with a small project. Therefore, the number of individuals with a large project increases while the number of individuals with a small or an intermediate project decreases, which is in accordance with the simulation results in fig. 6 (a) and (b) for $\xi=0.8$.

\subsection{\label{subsec:levelB} equilibrium between the frequencies of cooperators and the preferred project scales}
Suppose that the system has evolved to the equilibrium state where the cooperators and the defectors coexist. For a given group size $n$, the averaged payoff of cooperators is $\bar{P}_C=\frac{r\bar S_C}{n}-\frac{\bar S_C}{nf_C}$ and the averaged payoff of defectors is $\bar{P}_D=\frac{r\bar S_D}{n}$. In the equilibrium state, the averaged payoffs of cooperators and defectors should be satisfied with the equation $\bar{P}_C=\bar{P}_D$, that is,
\begin{equation}
\label{eq.4}
\frac{r\bar S_C}{n}-\frac{\bar S_C}{nf_C}=\frac{r\bar S_D}{n},
\end{equation}
which indicates that the cooperators tend to choose a larger project than the defectors. Consider a typical case $\bar S_C=\bar S_D+\varepsilon$, we obtain
\begin{equation}
\label{eq.4}
\frac{r\bar S_C}{n}-\frac{\bar S_C}{nf_C}=\frac{r(\bar S_C-\varepsilon)}{n}.
\end{equation}
The relationship between the frequencies of cooperators and the project scales selected by the cooperators becomes
\begin{equation}
\label{eq.5}
f_C=\frac{\bar S_C}{r\varepsilon}.
\end{equation}

The above equation indicates that, for a given r, the frequencies of cooperators are related to the project scales selected by the cooperators and the difference between the project scales selected by the cooperators and the defectors. The larger the preferred project scale $S_C$, the larger the frequency of cooperators $f_C$. The smaller the difference between $S_C$ and $S_D$, the larger the frequency of cooperators $f_C$. Such results are in accordance with the simulation results in fig. 3 and fig. 5.

\section{Summary}
\label{sec:summary}
By incorporating project selection mechanism into the threshold public goods game, we have investigated the coupling effects of mutation and imitation in updating one's preferred project scale on the evolution of cooperation. As there are more mutants than imitators, the evolution of cooperation is governed by the mutation process, which leads to the suppression of cooperation. As there are more imitators than mutants, the evolution of cooperation is governed by the coupling of the mutation and imitation processes, which leads to the widespread of cooperation. The suppression or promotion of cooperation in the present model is related to the coevolving of individual strategies and preferred project scales. A higher level of cooperation makes the individuals selecting a large project earn more, which attracts more individuals to select large projects and the average value of the preferred project scales increases. An increase in the preferred project scales makes the individuals with cooperation strategy earn more, which attracts more individuals to adopt cooperation strategy and the frequency of cooperators increases. A lower level of cooperation makes the individuals selecting a small project earn a positive return, which attracts more individuals to select small projects and the average value of the preferred project scales decreases. A decrease in the preferred project scales makes the individuals with defection strategy earn more, which attracts more individuals to adopt defection strategy and the frequency of cooperators decreases.

In the present work, we only investigated the role of project selection mechanism in the evolution of cooperation for a homogeneous population. Whether or not the project selection mechanism is beneficial for a heterogeneous population is another challenging issue that deserves further investigation.

\section*{Acknowledgments}
This work is the research fruits of Humanities and Social Sciences Fund sponsored by Ministry of Education of the People's Republic of China (Grant Nos. 19YJAZH120, 17YJAZH067, 18YJA790115, 17YJC790149) and National Natural Science Foundation of China (Grant Nos. 71371165, 11865009, 71471161, 71631005, 71773105, 71471031, 71772030, 71702025), Zhejiang Provincial Natural Science Foundation of China (Grant No. LY17G030024), the Program of Distinguished Professor in Liaoning Province (Grant No. [2018]35).





\bibliographystyle{model1-num-names}



\end{document}